\documentclass[twocolumn]{aastex631}

\usepackage{rotating}

\usepackage{graphicx}

\makeatletter
\let\jnl@style=\rm
\def\ref@jnl#1{{\jnl@style#1}}

\def\aj{\ref@jnl{AJ}}                   
\def\actaa{\ref@jnl{Acta Astron.}}      
\def\araa{\ref@jnl{ARA\&A}}             
\def\apj{\ref@jnl{ApJ}}                 
\def\apjl{\ref@jnl{ApJ}}                
\def\apjs{\ref@jnl{ApJS}}               
\def\ao{\ref@jnl{Appl.~Opt.}}           
\def\apss{\ref@jnl{Ap\&SS}}             
\def\aap{\ref@jnl{A\&A}}                
\def\aapr{\ref@jnl{A\&A~Rev.}}          
\def\aaps{\ref@jnl{A\&AS}}              
\def\azh{\ref@jnl{AZh}}                 
\def\baas{\ref@jnl{BAAS}}               
\def\bac{\ref@jnl{Bull. astr. Inst. Czechosl.}}
\def\caa{\ref@jnl{Chinese Astron. Astrophys.}}
\def\cjaa{\ref@jnl{Chinese J. Astron. Astrophys.}}
\def\icarus{\ref@jnl{Icarus}}           
\def\jcap{\ref@jnl{J. Cosmology Astropart. Phys.}}
\def\jrasc{\ref@jnl{JRASC}}             
\def\memras{\ref@jnl{MmRAS}}            
\def\mnras{\ref@jnl{MNRAS}}             
\def\na{\ref@jnl{New A}}                
\def\nar{\ref@jnl{New A Rev.}}          
\def\pra{\ref@jnl{Phys.~Rev.~A}}        
\def\prb{\ref@jnl{Phys.~Rev.~B}}        
\def\prc{\ref@jnl{Phys.~Rev.~C}}        
\def\prd{\ref@jnl{Phys.~Rev.~D}}        
\def\pre{\ref@jnl{Phys.~Rev.~E}}        
\def\prl{\ref@jnl{Phys.~Rev.~Lett.}}    
\def\pasa{\ref@jnl{PASA}}               
\def\pasp{\ref@jnl{PASP}}               
\def\pasj{\ref@jnl{PASJ}}               
\def\rmxaa{\ref@jnl{Rev. Mexicana Astron. Astrofis.}}%
\def\qjras{\ref@jnl{QJRAS}}             
\def\skytel{\ref@jnl{S\&T}}             
\def\solphys{\ref@jnl{Sol.~Phys.}}      
\def\sovast{\ref@jnl{Soviet~Ast.}}      
\def\ssr{\ref@jnl{Space~Sci.~Rev.}}     
\def\zap{\ref@jnl{ZAp}}                 
\def\nat{\ref@jnl{Nature}}              
\def\iaucirc{\ref@jnl{IAU~Circ.}}       
\def\aplett{\ref@jnl{Astrophys.~Lett.}} 
\def\apspr{\ref@jnl{Astrophys.~Space~Phys.~Res.}}
\def\bain{\ref@jnl{Bull.~Astron.~Inst.~Netherlands}} 
\def\fcp{\ref@jnl{Fund.~Cosmic~Phys.}}  
\def\gca{\ref@jnl{Geochim.~Cosmochim.~Acta}}   
\def\grl{\ref@jnl{Geophys.~Res.~Lett.}} 
\def\jcp{\ref@jnl{J.~Chem.~Phys.}}      
\def\jgr{\ref@jnl{J.~Geophys.~Res.}}    
\def\jqsrt{\ref@jnl{J.~Quant.~Spec.~Radiat.~Transf.}}
\def\memsai{\ref@jnl{Mem.~Soc.~Astron.~Italiana}}
\def\nphysa{\ref@jnl{Nucl.~Phys.~A}}   
\def\physrep{\ref@jnl{Phys.~Rep.}}   
\def\physscr{\ref@jnl{Phys.~Scr}}   
\def\planss{\ref@jnl{Planet.~Space~Sci.}}   
\def\procspie{\ref@jnl{Proc.~SPIE}}   

\makeatother

\newcommand{\osoeight}{\,\emph{OSO$-$8}}
\newcommand{\xmm}{\,\emph{XMM}$-$Newton}

\newcommand{\nus}{\,\emph{NuSTAR}}
\newcommand{\astrosat}{\,\emph{AstroSat}}

\newcommand{\ixpe}{\,\emph{IXPE}}

\newcommand{\chandra}{\,\emph{Chandra}}
\newcommand{\rxte}{\,\emph{RXTE}}

\newcommand{\asca}{\,\emph{ASCA}}
\newcommand{\ginga}{\,\emph{GINGA}}
\newcommand{\xrism}{\,\emph{XRISM}}

\newcommand{\maxi}{\,\emph{MAXI}}
\newcommand{\gsc}{\,MAXI/\emph{GSC}}

\newcommand{\suzaku}{\,\emph{Suzaku}}

\newcommand{\msun}{$M_\odot$}
\newcommand{\rsun}{$R_\odot$}
\newcommand{\xray}{X-ray}
\newcommand{\kalpha}{K$\alpha$}
\newcommand{\Healpha}{He$\alpha$}
\newcommand{\Lyalpha}{Ly$\alpha$}
\newcommand{\kbeta}{K$\beta$}

\renewcommand{\arraystretch}{1.8}

\newcommand{\src}{\,Cen X--3}
\newcommand{\source}{\,\emph{Centaurus X--3}}

\usepackage{hyperref}
\hypersetup{linkcolor=red,citecolor=blue,filecolor=cyan,urlcolor=magenta}

\begin{document}

\title{XMM-Newton view of pulsating iron fluorescent emission from Centaurus X-3}

\author[0000-0002-7391-5776]{Kinjal Roy}
\affiliation{Raman Research Institute, C. V. Raman Avenue, Sadashivanagar, Bengaluru - 560 080, India.}

\author[0000-0001-9404-1601]{Hemanth Manikantan}
\affiliation{Raman Research Institute, C. V. Raman Avenue, Sadashivanagar, Bengaluru - 560 080, India.}
\affiliation{INAF – Istituto di Astrofisica e Planetologia Spaziali, Via del Fosso del Cavaliere, 100, 00133 Roma, Italy}

\author[0000-0002-8775-5945]{Biswajit Paul}
\affiliation{Raman Research Institute, C. V. Raman Avenue, Sadashivanagar, Bengaluru - 560 080, India.}



\begin{abstract}

\src\ is a bright, high-mass \xray\ binary pulsar. We present a pulse phase-resolved \xray\ spectral analysis of data from an archival \xmm\ observation of \src\ taken during the high state of the source. The observation was entirely in the out-of-eclipse part of the binary orbit. We study the pulse phase variability of the three \kalpha\ fluorescent emission lines from near-neutral, Helium-like and Hydrogen-like iron, along with the iron \kbeta\ emission line. All four lines show clear modulation with the pulse phase of the neutron star, and modulation is found to be higher for the lines from highly ionised iron compared to the neutral lines. Structures within the light travel distance corresponding to the pulse period of the neutron star are likely responsible for the pulse phase modulation of the emission lines. We have also investigated the orbital phase dependence of the pulse phase variability in the iron lines by dividing the data into four segments at different orbital phases of \src. The pulse phase modulation behaviour of the four lines is quite identical at different orbital phases of \src\, indicating the pulsed iron emission region is persistent in nature and probably phase aligned with respect to the observer. The accretion stream intercepting the line of sight can probably produce the observed phase dependence of the iron fluorescence emission lines.


\end{abstract}




\section{Introduction} \label{sec:intro}

\source\ (henceforth \src) is a High Mass \xray\ Binary (HMXB) system which was discovered with a rocket-borne proportional counter in 1967~\citep{Chodil_cenx3_discovery_1967}. The compact object in the HMXB was determined to be a Neutron Star (NS) when periodicity ($P_{spin}$) of $\sim4.8$ s was detected from the \xray\ data~\citep{Giacconi_cenx3_pulsations_1971}, making it one of the first detected \xray\ pulsars. The system is an eclipsing binary with an almost circular orbit~\citep{Schreier_CenX3_ecllipse_1972, CenX3_Raichur_2010, Klawin_CenX3_ephemeris_2023}.
The companion star, V779 Cen, was identified as an O6-8 III star with a mass of $\sim 21$ \msun\ with a radius of $\sim 12$ \rsun~\citep{Hutchings_CenX3_companion_radius_1979}.
The NS mass was estimated to be $ 1.34^{+0.16}_{-0.14}$ \msun\ from optical spectroscopic data~\citep{van_der_meer_cenx3_2007}. 
The binary has an orbital period ($P_{orb}$) of about 2.09 days with a period decay rate of ($\dot{P}_{orb}$/$P_{orb}$)  approximately $-1.8 \times 10^{-6}$ yr$^{-1}$~\citep{CenX3_Raichur_2010, Falanga_cenx3__2015, Klawin_CenX3_ephemeris_2023}

A cyclotron line has been observed in the source spectrum near $\sim$ 30 keV, giving a magnetic field strength of about $3 \times 10^{12}$ Gauss~\citep{Nagase_cenx3_ginga_1992, Gunjan_NuStart_CRSF_2021}. Spin phase resolved analysis of the same \nus\ study showed a positive correlation of the cyclotron line energy with source flux~\citep{Tamba_CenX3_variability_2023}. 
From the stable pulse profile across the entire observation the study also concluded the presence of a stable accretion stream in the binary system.
Subsequent study of the cyclotron line with \nus\ over two orbits showed that there were no clear correlation between luminosity of the source and the cyclotron line energy, with a possible critical luminosity lying in the range $0.5 - 4 \times 10^{37}$ erg/s~\citep{CenX3_NuStar_Liu_2024}.
Polarisation measurements from the source with \ixpe\ show no detected polarisation in the low-intensity level, while the brighter observation gives a very low degree of polarisation of $\sim 6$ percent~\citep{Tsygankov_cenx3_ixpe_2022}.
The low polarisation signal was explained by a combination of multiple factors, including overheated upper atmosphere from the NS and mixing of polarisation signals from the two magnetic poles.

\src\ is one of the most studied HMXB systems. The luminosity of \src\ varies by almost a factor of 40, with the maximum luminosity reaching $5 \times 10^{37}$ erg/s, which indicates that the predominant mode of accretion is probably via a disc~\citep{Suchy_cenx3_rxte_2008, Raichur_CenX3_long_term_intensity_2008}.
Orbit averaged long term light curve shows strong intensity variations and a period of 120 - 165 days was noticed in earlier observations~\citep{Priedhorsky_1983}.
A super-orbital period of 220 days was proposed with 6 years of \maxi\ data, which was attributed to the precession of the accretion disc~\citep{Torregrosa_CenX3_MAXI_2022}. However, subsequent analysis of longer duration data of \src\ from \maxi\ has not shown any persistent long-term periodicity~\citep{Ajith_cenx3_2024}. Investigations of the long-term light curves have further revealed that the eclipse ingresses and egresses were shorter and sharper in high-intensity levels compared to low-intensity levels, indicative of varying disc obscuration~\citep{Raichur_CenX3_long_term_intensity_2008, Jincy_Paul_cenX3_2010}. A tentative link between torque reversal and orbital profile was also suggested from long-term observations~\citep{Liao_cenx3_torque_orbital_profile_2024}. In a further study of the orbital characteristics with long-term intensity variations with \gsc, ~\cite{Ajith_cenx3_2024} found an increase in the hardness ratio in the orbital phases between 0.5 and 0.9 in the lowest intensity level, which can be interpreted as due to the presence of a wake-like structure.

An iron emission line was first discovered in the source spectrum from \osoeight\ observation~\citep{White_accreting_pulsar_review_1983}. The iron line was later found to be a combination of emission lines from multiple ionisation states of iron, namely neutral and near-neutral iron (henceforth denoted as Fe I), Helium-like iron (Fe XXV) and Hydrogen-like iron (Fe XXVI) species from \asca\ data~\citep{Ebisawa_cenx3_asca_1996}. The presence of the three iron lines has been used to probe different regions of the reprocessing environment as they originate from different regions in the binary environment~\citep{Ebisawa_cenx3_asca_1996, Naik_cenx3_xmm_2012}. 

Even though iron fluorescence emission is ubiquitous in \xray\ binaries~\citep{XMM_Fe_k_alpha_study}, pulsations in the iron line have only been observed in a few sources like \src~\citep{Day_CenX3_pulsed_Fe_1993, Tamba_CenX3_variability_2023}, Her X$-$1~\citep{Choi_HerX1_Fe_pulsation_1994, Vasco_HerX1_pulsating_iron_line_2013}, V0332+53~\citep{Bykov_V0332p53_pulsating_iron_line_2021}, and GX 301$-$2~\citep{Roy_GX301m2_2024}. Other emission lines have also been observed to be variable with NS rotation, like the Ne IX \kalpha\ fluorescent line in the ultra-compact \xray\ binary 4U 1626$-$67~\citep{Beri_4U_1626m67_emission_line_pulsation_2018}.

Pulse phase dependent modulation in the iron \kalpha\ emission line was discovered in \src\ with \ginga\ data~\citep{Day_CenX3_pulsed_Fe_1993} and later confirmed with \nus\ data~\citep{Tamba_CenX3_variability_2023}. However, for both these cases, the entire iron \kalpha\ complex was modelled with a single Gaussian line. Iron line variations with spin of the NS can be produced if there is an anisotropic distribution of matter in the line-producing region and if the light travel time across the region is smaller than the pulse period of the NS. For the case of slow rotating pulsars ($P_{spin}$ $\gtrsim$ 100 s) like GX 301$-$2 or Vela X$-$1 the light travel distance corresponding to the pulse period of the NS is larger than the binary orbit, therefore any anisotropies in the wider reprocessing environment can give rise to pulse phase dependency of the iron emission lines. For fast spinning pulsars ($P_{spin}$ $\lesssim$ 10 s) like Her X$-$1, 4U 1626$-$67 and \src, the pulse phase dependence of fluorescent lines can only be produced close to the NS. In \src, the binary orbit size ($\sim$ 40 lt-s) is much larger than the light travel distance corresponding to the pulse period of the NS ($\sim$ 4.8 lt-s), making it possible to study the anisotropies in the line-producing region in the inner orbital region in the vicinity of the NS. 

The large effective area and good spectral resolution of \xmm\ allow us to resolve the iron fluorescence emission complex and study the variation in the individual iron lines. We take advantage of the excellent spectral and timing properties of \xmm\ to investigate the pulse phase dependence of the individual iron emission lines from an archival observation of \src\ taken in 2006. We structure the manuscript as follows, in Section~\ref{sec:inst} we introduce the details of \xmm\ instrument, the observation details and the data reduction techniques. In Section~\ref{sec:analysis}, we describe the data analysis methods and the results. In Section~\ref{sec:disc}, we discuss the implications of the results and summarise the study in Section~\ref{sec:summary}.


\section{Instruments, observation and data reduction} \label{sec:inst}

\begin{figure*}
    \centering
    \includegraphics[width=0.99\linewidth]{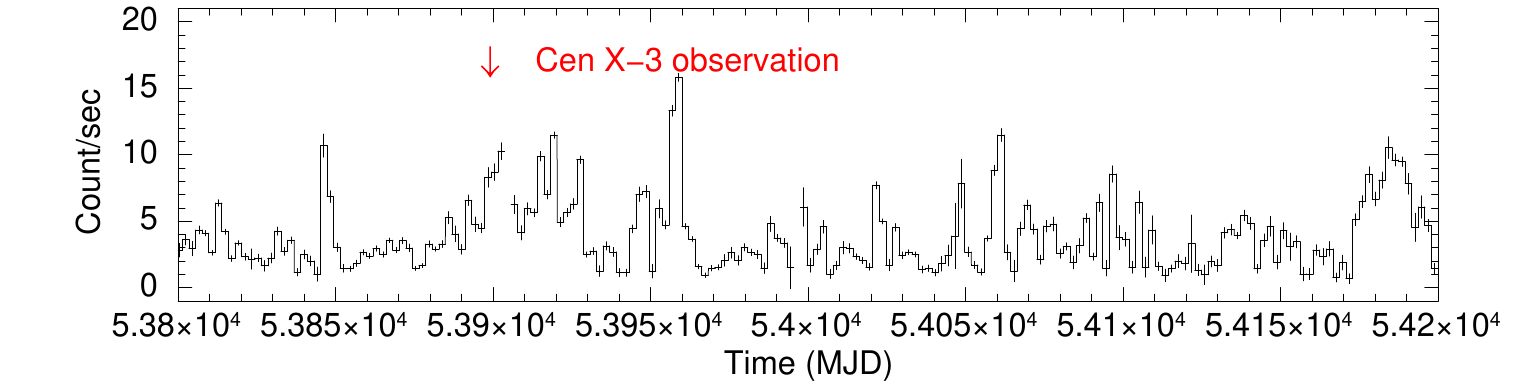}
    \caption{The long-term lightcurve of \src\ with \rxte/ASM is plotted in black with a bin size corresponding to the orbital period of the source. The red marker corresponds to the \xmm\ observation of the source.}
    \label{fig:rxte_asm_long_term}
\end{figure*}

X-ray Multi-Mirror Mission (\xmm)~\citep{XMM_Newton_observatory}, an \xray\ observatory of the European Space Agency (ESA), consists of three \xray\ detectors named \textit{European Photo Imaging Camera} (EPIC) at the focal plane of three \xray\ telescopes. Two of the EPIC cameras have MOS CCD detectors~\citep{XMM_MOS_paper} and the third contains a PN CCD detector~\citep{XMM_PN_paper}. There are two additional \textit{Reflection Grating Spectrometer} (RGS) telescopes for high-resolution \xray\ spectroscopy~\citep{XMM_RGS} and an \textit{Optical Monitor} (OM) for optical and UV imaging~\citep{XMM_OM}.

We present the results from analysis of one of the longest observations of \src\ with \xmm\ taken on 12th August 2006 (MJD 53898.938 - MJD 53899.860).
The observation (ObsID: 0400550201), lasting $\sim$ 81 ks, was taken in the \textit{TIMING MODE} with the MOS2 and PN detectors. The EPIC cameras operate in the energy range of 0.5$-$10.0 keV. We have utilized the EPIC$-$PN data for this work due to its large effective area.

We used the Science Analysis System (SAS) version 19.1.0 for all data extractions. We followed the instructions provided in the SAS \textit{Data Analysis Threads}\footnote{\url{https://www.cosmos.esa.int/web/xmm-newton/sas-threads}}. The data reduction procedures relevant for \textit{TIMING MODE} were followed from the \textit{XMM-Newton ABC Guide}\footnote{\url{https://heasarc.gsfc.nasa.gov/docs/xmm/abc/abc.html}}. We have used the latest calibration files to reduce the data. 

We have only considered events with \texttt{FLAG = 0} and with  \texttt{PATTERN <= 4} for the PN data\footnote{\url{https://www.cosmos.esa.int/web/xmm-newton/sas-threads}}. The source region was extracted from columns centred around the brightest zone of the image (\texttt{RAWX IN [28:44]}), and the background was selected from columns away from the source region (\texttt{RAWX IN [5:15]}). Solar system barycentre correction was performed on the event file by SAS tool \texttt{barycen}. The photon time of arrivals were corrected for the motion of NS in the binary orbit using orbital ephemeris from~\citet{Klawin_CenX3_ephemeris_2023}. The source and background light curves with a resolution of 0.0001 s and spectra were generated from the aforementioned regions. The response and ancillary files were generated using the \texttt{rmfgen} and \texttt{arfgen} tools.  We checked for pile-up in the data using the SAS tool \texttt{epatplot} and found no noticeable pile-up.

The long-term lightcurve from \rxte/\textit{ASM} binned with the orbital period of \src\ is shown in Fig.~\ref{fig:rxte_asm_long_term}. The epoch of the \xmm\ observation is shown in red in the same.

\section{Data analysis} \label{sec:analysis}

\subsection{Timing analysis} \label{sec:timing_analysis}

\begin{figure}
    \centering
    \includegraphics[width=0.99\linewidth]{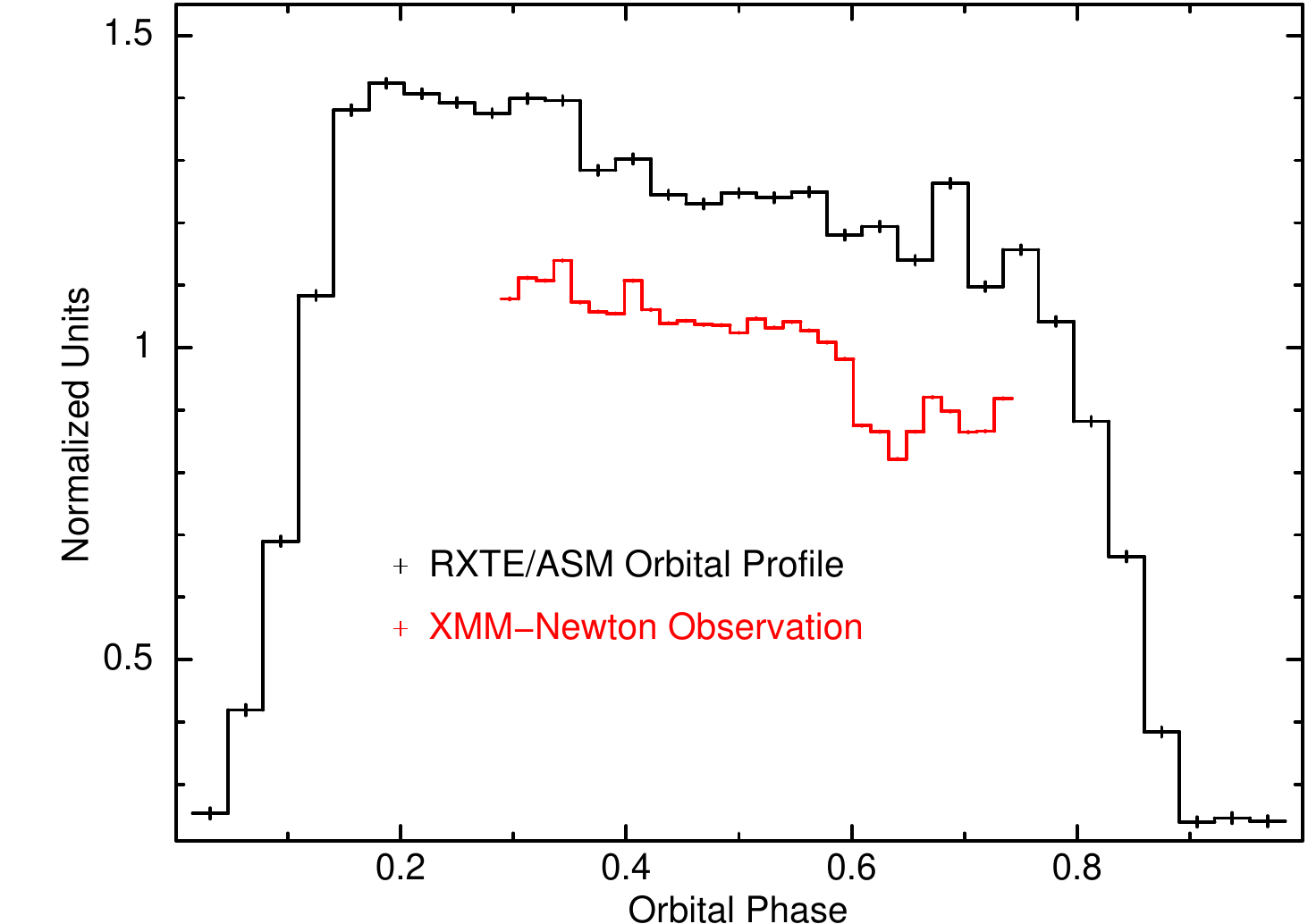}
    \caption{The long term orbital profile of \src\ is plotted in black from \rxte/ASM in the $2-10$ keV energy range normalised by the average count rate. The \xmm\ observation is plotted in red normalised by the average count rate.}
    \label{fig:cenx3_rxte_orbital_profile}
\end{figure}

\src\ is an eclipsing binary source with a high inclination of around $79^{\circ}$~\citep{Sanjurfo_ferrin_CenX3_XMM_2021}.
The orbital \xray\ intensity is plotted in Fig.~\ref{fig:cenx3_rxte_orbital_profile} from \rxte/ASM in the $2-10$ keV energy range using the orbital ephemeris from \citet{Klawin_CenX3_ephemeris_2023}. The \xmm\ observation was taken during the out-of-eclipse phase of the orbit between the orbital phase $0.297-0.738$, where zero phase is assigned to mid-eclipse.

\begin{figure}
    \centering
    \includegraphics[width=0.99\linewidth]{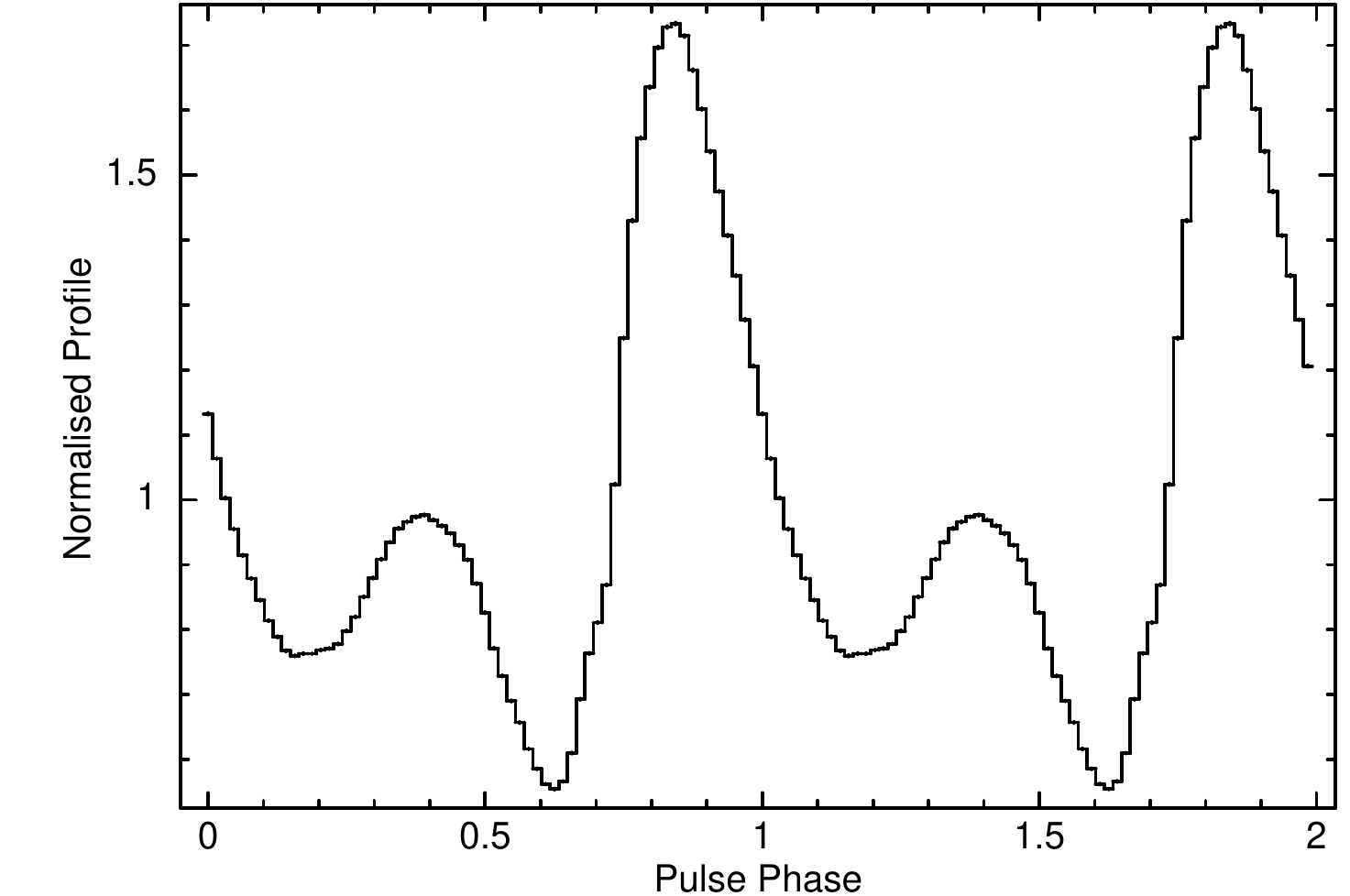}
    \caption{The pulse profile obtained by folding the 0.5-10 keV \xmm\ EPIC$-$PN lightcurve with the pulse period of the pulsar of 4.805069 s normalised by the average count rate of the source.}
    \label{fig:pulse_profile}
\end{figure}

The light curve shows variability at multiple timescales along with pulsations at $\sim4.8$ s. The pulse period of the source was calculated to be 4.805069 (1) s from the barycentre and binary orbit corrected data using the epoch folding tool \texttt{efsearch}. The average pulse profile from the entire observation obtained by folding the lightcurve is plotted in Fig~\ref{fig:pulse_profile}. The pulse profile of \src\ in the 0.5$-$10 keV energy range is double-humped in nature with the primary and secondary peak having a phase separation of $\sim0.45$. The pulse fraction (PF) of the 0.5$-$10 keV lightcurve is 51.5$\pm$0.2 \%, where the PF for a pulse profile (P) is defined as, PF = ($P_{max}$ - $P_{min}$)/($P_{max}$ + $P_{min}$).

\subsection{Time-averaged spectroscopy} \label{sec:spectral_analysis}

The spectrum of the source was fitted using the \texttt{HEASoft} spectral fitting tool XSPEC version 12.12.0~\citep{XSPEC}. The absorption by interstellar medium was modelled using the Tuebingen-Boulder absorption model \texttt{TBabs} with abundance taken from~\citet{Wilm_abund} and the photoelectric cross sections from~\citet{Vern_cs}. The spectra used in this study have been optimally binned based on~\citet{Optimal_binning_Kaastra_Bleeker_2016}.  For very bright \xray\ sources, the spectra from \xmm\ EPIC$-$PN cameras can suffer from pileup effects, but no clear pile-up was observed in this source. The PN spectrum in timing mode has some energy calibration uncertainties~\footnote{\url{https://xmmweb.esac.esa.int/docs/documents/CAL-TN-0083.pdf}} that are more prominent at lower energies, and since the main purpose of the current study is to investigate the iron line variations, we only consider the spectra in a limited energy band of $5.0-9.0$ keV around the iron fluorescent line. We fitted the $5.0-9.0$ keV EPIC$-$PN spectrum of \src\ to study the iron fluorescent emission. 

\begin{figure}
    \centering
    \includegraphics[width=0.99\linewidth]{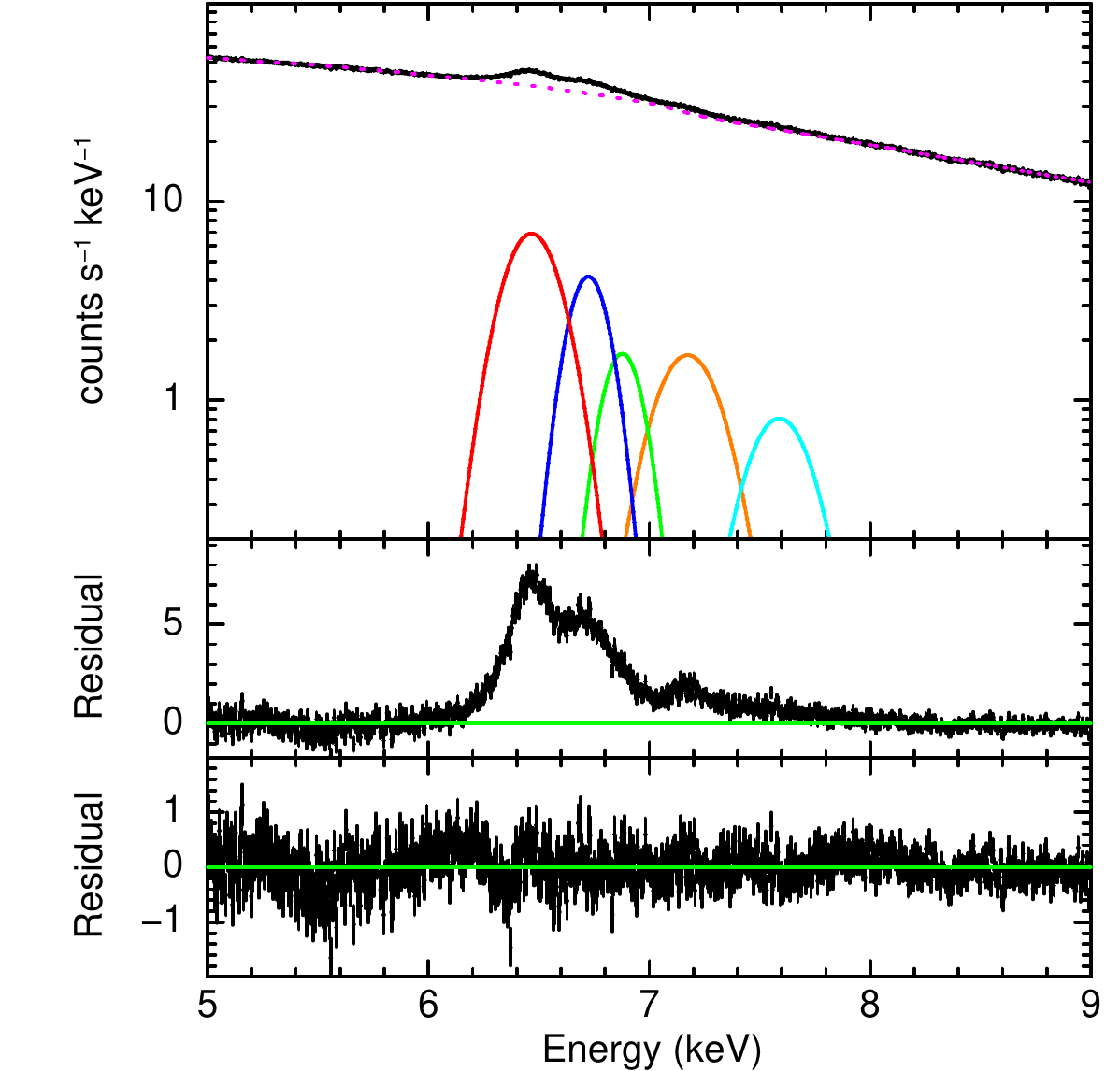}
    \caption{The $5.0-9.0$ keV \xmm\ EPIC$-$PN spectrum of \src\ is shown in the figure. \textit{Top Panel} contains the data and the best-fit model components. 
    \textit{Middle Panel} show the residuals when the strength of the emission lines are all set to zero.
    \textit{Bottom Panel} show the residuals between the data and the best-fit model.}
    \label{fig:overall_sp}
\end{figure}

The \xray\ continuum was modelled with an absorbed power-law component. Multiple emission lines were present on top of the continuum emission. In total, we detected five emission lines in the aforementioned energy range, the most prominent being the 6.4 keV neutral iron \kalpha\ line. The final spectral model used is \texttt{tbabs} $\times$ ( \texttt{powerlaw} $+$ 5 $\times$ \texttt{Gaussians} ). The fluorescence emission from highly ionised Fe XXV and Fe XXVI species was also present, along with iron \kbeta\ and nickel \kalpha\ emission lines. The best-fit spectral model and the data are plotted in Fig.~\ref{fig:overall_sp}. The emission lines from near-neutral iron \kalpha\ (red), iron XXV (blue), iron XXVI (green), iron \kbeta\ (orange), and nickel \kalpha\ (cyan) could be resolved from EPIC$-$PN spectrum~\citep{Sanjurfo_ferrin_CenX3_XMM_2021} (Fig.~\ref{fig:overall_sp}). The line detection significance quantified as the ratio of the strength of the lines and the error on the estimate of the line strength for the Fe I \kalpha, Fe XXV \Healpha, Fe XXVI \Lyalpha\, Fe \kbeta\ and Ni \kalpha\ was calculated to be 68, 36, 15, 22 and 11 $\sigma$. The equivalent width of all the emission lines is in the range of $10-50$ eV. The best fit spectral parameters are given in Table.~\ref{tab:av_spectra} with errors quoted at 90 \% confidence level.

\begin{table}[hbt!]
\centering
\begin{tabular}{|c|c|c|c|}
\hline
Model                     & Parameter  & Units     & Best Fit Value          \\ \hline
Absorption                & $N_{H}$    & $10^{22}$ & $8.18^{+0.32}_{-0.32}$  \\ \hline
Power-law                 & $\Gamma$   &           & $1.27^{+0.01}_{-0.01}$  \\ \hline
                          & Norm$^{a}$ &           & $0.60^{+0.01}_{-0.01}$  \\ \hline
Fe K$\alpha$              & Norm$^{b}$ & $10^{-3}$ & $2.99^{+0.04}_{-0.04}$  \\ \hline
                          & Eqw        & eV        & $52.25^{+0.87}_{-0.94}$ \\ \hline
Fe XXV                    & Norm$^{b}$ & $10^{-3}$ & $1.36^{+0.04}_{-0.04}$  \\ \hline
                          & Eqw        & eV        & $23.80^{+0.67}_{-0.67}$ \\ \hline
Fe XXVI                   & Norm$^{b}$ & $10^{-4}$ & $5.85^{+0.40}_{-0.40}$  \\ \hline
                          & Eqw        & eV        & $10.75^{+0.77}_{-0.76}$ \\ \hline
Fe K$\beta$               & Norm$^{b}$ & $10^{-3}$ & $1.01^{+0.05}_{-0.05}$  \\ \hline
                          & Eqw        & eV        & $20.22^{+0.87}_{-0.85}$ \\ \hline
Ni K$\alpha$              & Norm$^{b}$ & $10^{-4}$ & $5.41^{+0.49}_{-0.49}$   \\ \hline
                          & Eqw        & eV        & $11.91^{+1.11}_{-1.03}$ \\ \hline
Flux$^{c}$                &            & $10^{-9}$ & $3.64^{+0.02}_{-0.02}$  \\ \hline
$\chi^{2}$ (d.o.f.)       &            &           & 1012.12 (791)           \\ \hline
\end{tabular}
\caption{The table contains the best fit spectral parameter corresponding to the phase-averaged spectrum of \src\ with errors quoted at 90 \% confidence interval. $^{a}$ The powerlaw normalization is in units of photons keV$^{-1}$ cm $^{-2}$ at 1 keV. $^{b}$ The units of the normalisation for the Gaussian profile is in the units of total number of photons s$^{-1}$ cm $^{-2}$ in the line. $^c$ The units for flux is erg s$^{-1}$ cm$^{-2}$. }
\label{tab:av_spectra}
\end{table}

\subsection{Pulse phase-resolved spectroscopy} \label{sec:PRS}

We performed pulse phase resolved spectroscopy of the entire observation to study the variability of the fluorescence lines with the spin of the NS using barycentre and orbital motion corrected event data. The pulse profile was divided into twenty equal phase bins, and the spectrum was extracted from each phase segment. The same spectral model we fitted to the phase-averaged spectrum was used to fit the spectrum from each of these data sets. 
The centres and widths of the emission lines were fixed to the best-fit values obtained from the phase-averaged spectrum due to limited photon statistics. We put emphasis on the three \kalpha\ lines from neutral, Helium-like and Hydrogen-like iron species and the iron \kbeta\ line, which are the most prominent.

\begin{figure}
    \centering
    \includegraphics[width=0.99\linewidth]{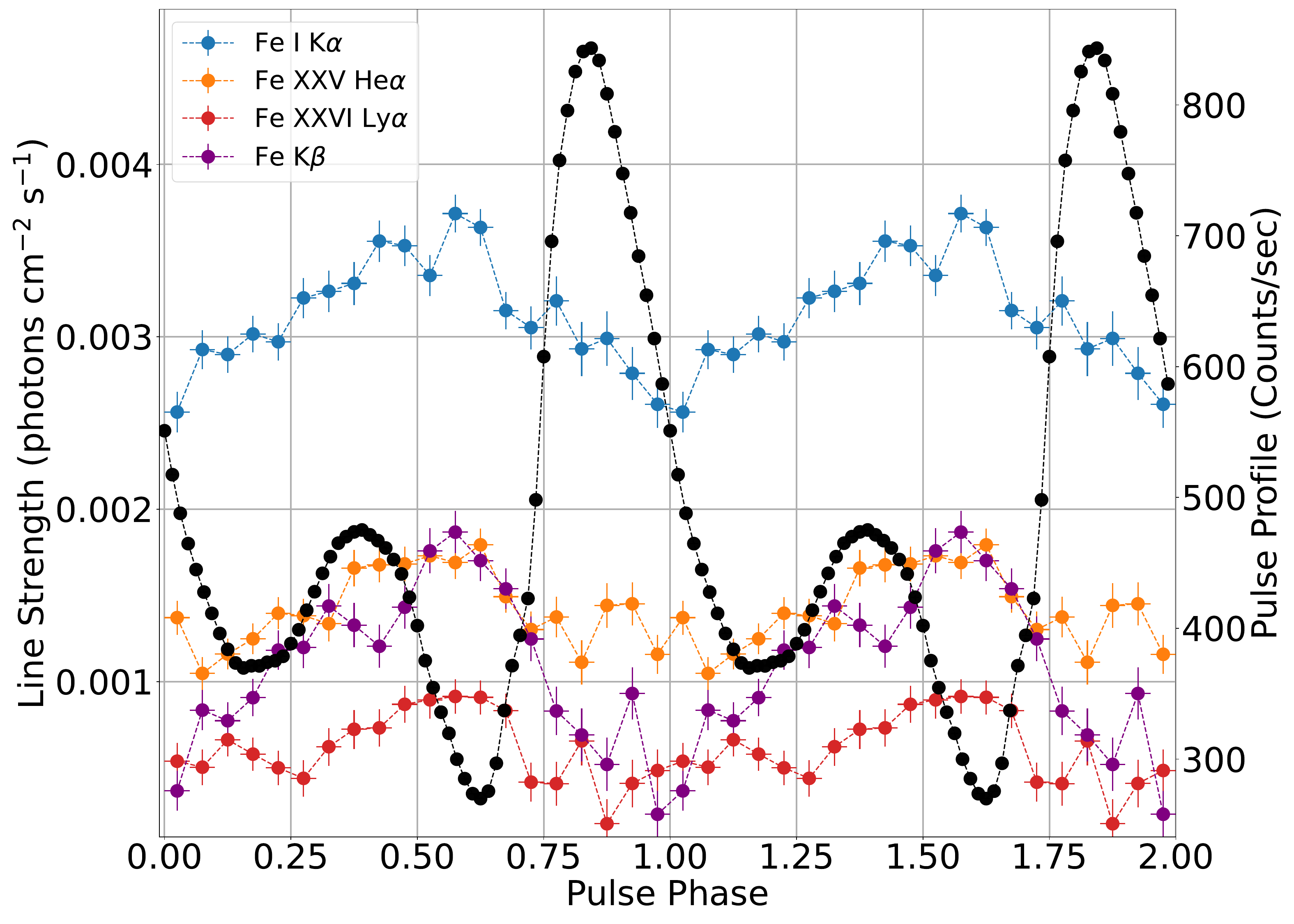}
    \caption{The variation of the four emission lines, from near neutral Fe \kalpha\ (blue), Fe XXV \Healpha\ (orange), Fe XXVI \Lyalpha\ (red) and Fe \kbeta\ (purple) is plotted with pulse phase of the NS along with the pulse profile in the $0.5-10$ keV range (black). }
    \label{fig:PRS_line_strength_pulse_profile}
\end{figure}

The emission line strength, i.e. the normalisation of the Gaussian profile used to model the emission line, is plotted in Fig.~\ref{fig:PRS_line_strength_pulse_profile} as a function of pulse phase.
The pulse profile in the $0.5-10$ keV range is plotted in black for reference. The Fe I \kalpha\ (blue), Fe XXV \Healpha\ (orange), Fe XXVI \Lyalpha\ (red), and Fe \kbeta\ (purple) all show variability with the pulse phase as seen in Fig.~\ref{fig:PRS_line_strength_pulse_profile}. The modulation in all four iron line intensities is quite similar, and the peak of the line emission is not aligned with the peak of the pulse profile. We discuss the results further in Section~\ref{sec:disc}.

\subsection{Variation of iron emission lines with different trial pulse periods} \label{sec:iron_variation}

\begin{figure}
    \centering
    \includegraphics[width=0.99\linewidth]{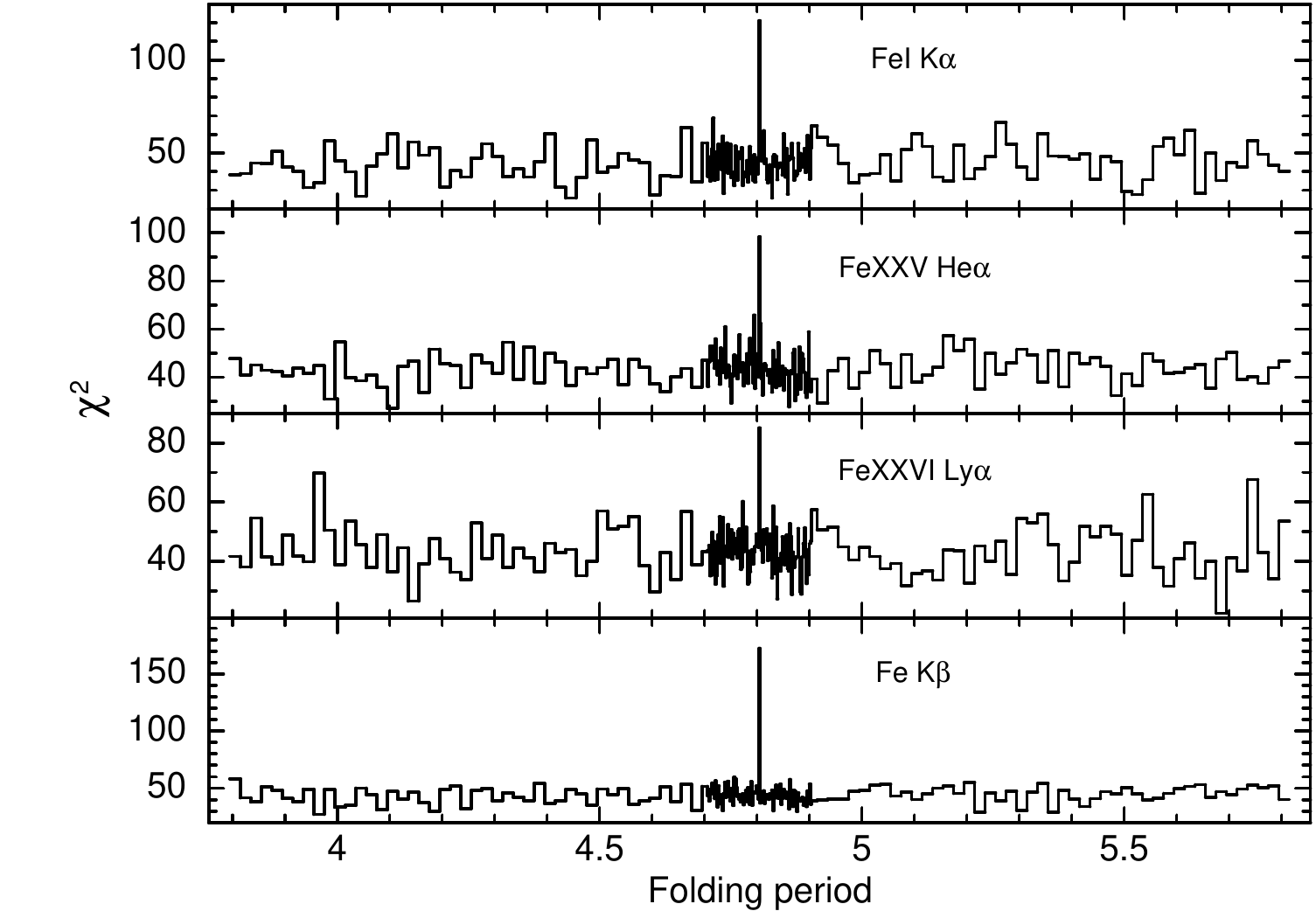}
    \caption{The figure shows the variation of the chi-square value of the pulse phase modulation of the iron line flux for twenty pulse phase bins with different trial pulse periods. The results for Fe I \kalpha, Fe XXV \Healpha, Fe XXVI \Lyalpha\ and Fe \kbeta\ are plotted from top in different panels.}
    \label{fig:iron_chi_sq}
\end{figure}

In order to confirm the presence of modulations in the iron emission lines with the pulse period of the NS, we performed pulse phase-resolved spectroscopy with multiple trial periods on either side of the NS pulse period of 4.805069 s. A trial period range of $\pm$0.1 s around the pulse period was investigated with a period resolution of 0.002 s, and a range of $\pm$1.0 s was investigated with a period resolution of 0.02 s. For each trial period, we divided the data into twenty equal phase bins and fitted the same spectral model used in the phase-averaged analysis (Section~\ref{sec:spectral_analysis}) to estimate the iron line strength in each phase interval. For each of the four iron line components, the twenty measured line flux values were fitted to a constant and the chi-square was determined for the best fit. This chi-square acts as a proxy for the strength of modulation present in the line emission for each trial period. The resulting chi-square distribution for near-neutral Fe I \kalpha, Fe XXV \Healpha, Fe XXVI \Lyalpha, and Fe \kbeta\ are shown against the trial periods in Fig.~\ref{fig:iron_chi_sq}.

The four iron emission lines clearly show a peak in chi-square only at the NS pulse period of 4.805069 s (Fig.~\ref{fig:iron_chi_sq}). Therefore, we can establish from this independent analysis that the variation of the iron fluorescent emission lines with rotation of the NS found in Section~\ref{sec:PRS} is an actual signal with a significant detection.

\subsection{Orbital-phase dependent pulse phase-resolved spectroscopy} \label{sec:TR_PRS}

\begin{figure}
    \centering
    \includegraphics[width=0.99\linewidth]{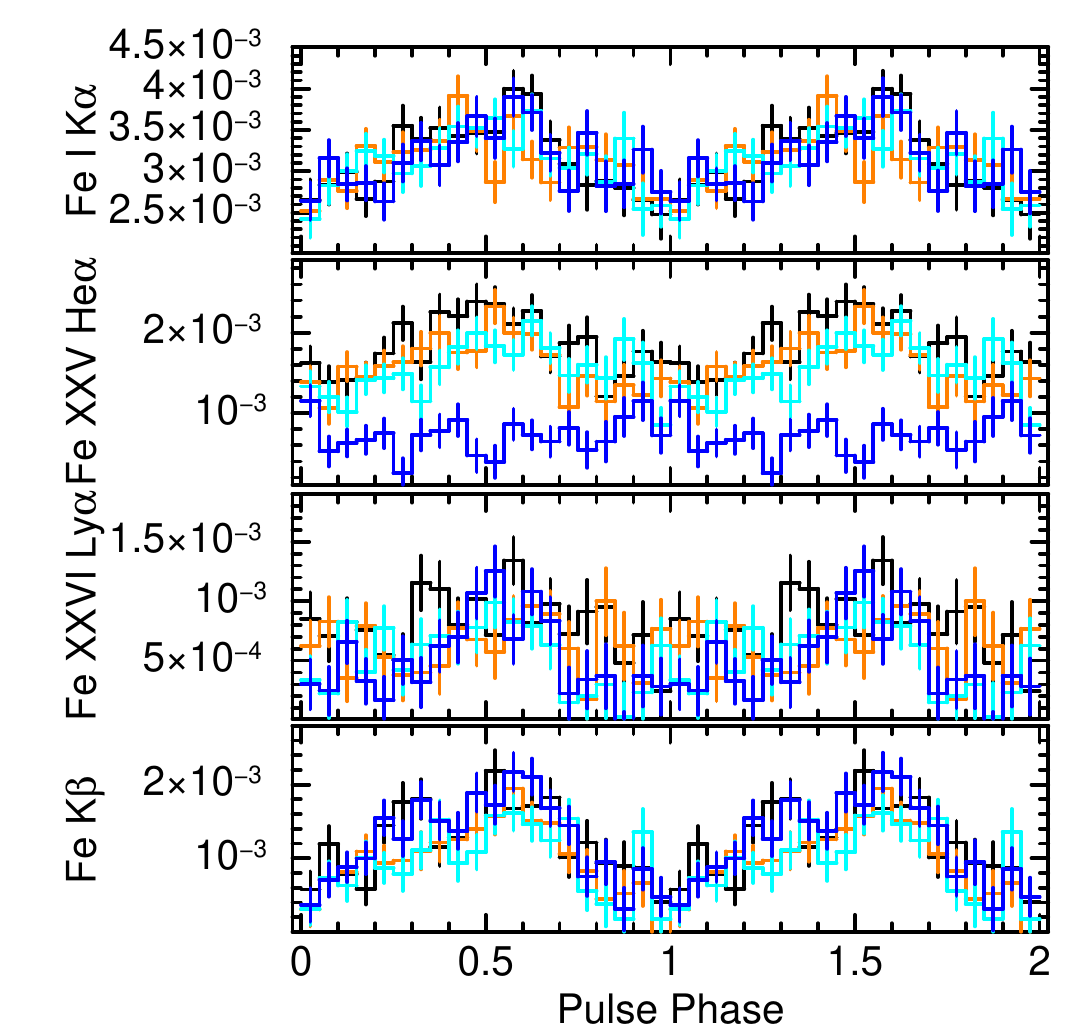}
    \caption{The figure shows the variation of the pulse phase modulation of the iron line fluxes as a function of orbital phase. From top, the results for Fe I \kalpha, Fe XXV \Healpha, Fe XXVI  \Lyalpha, and Fe \kbeta\ are plotted in different panels. The results from four different orbital phase segments are plotted in black ($0.297-0.407$), orange ($0.407-0.518$), cyan ($0.518-0.628$) and blue ($0.628-0.738$) colours.}
    \label{fig:TR_PRS}
\end{figure}

Pulsating iron line emission can originate from various parts of the binary environment that are smaller in size than the light travel distance for the pulse period of the NS. Any line forming region in the binary system whose position with respect to our line of sight to the NS changes with orbital phase is expected to show a variable phase difference between line flux modulation and the total pulse profile in different orbital phases. To evaluate this, we divided the total observation into four orbital phase segments. In each of the four segments, we probed the pulse phase dependence of the iron line emission similar to Section~\ref{sec:PRS}. The results for near-neutral Fe \kalpha, Fe XXV \Healpha, Fe XXVI \Lyalpha\ and Fe \kbeta\ are plotted in different panels in Fig.~\ref{fig:TR_PRS} in different colours for the different orbital phase segments. The orbital segments span approximately from $0.297-0.407$, $0.407-0.518$, $0.518-0.628$, and $0.628-0.738$ and are plotted in Fig.~\ref{fig:TR_PRS} in black, orange, cyan and blue, respectively. The near-neutral iron \kalpha\ line shows clear pulsations in all the segments. The most interesting observation is that the minima of the near-neutral iron \kalpha\ line strength occur at a similar pulse phase for each of the orbital phase segments. A similar pattern is also seen in the iron \kbeta\ emission line. In segments where pulsations have been detected, the modulation of the Helium-like and Hydrogen-like iron emission shows minima at a similar pulse phase. The strength of the Fe XXV \Healpha\ line in the fourth segment was low compared to the other segments, and no pulsations were detected in the Fe XXV \Healpha\ line. This is in agreement with time-resolved analysis with the \xmm\ data~\citep{Sanjurfo_ferrin_CenX3_XMM_2021}. Pulse phase modulation was not clearly detected in the second segment for the Fe XXVI \Lyalpha\ line.

\section{Discussion} \label{sec:disc}

In this manuscript, we report the results from analysis of data from an \xmm\ observation of \src\ taken in the out-of-eclipse part of its binary orbit. \src\ was observed for $\sim$ 81 ks between the orbital phase of $\sim$ 0.297 to 0.738. Pulsations were detected in the source with a period of 4.805069 s. The $5.0-9.0$ keV spectrum was modelled with an absorbed power-law continuum with multiple emission lines from iron and nickel \kalpha\ line. The detection significance of the Fe I \kalpha, Fe XXV \Healpha, Fe XXVI \Lyalpha\, Fe \kbeta\ and Ni \kalpha\ was calculated to be 68, 36, 15, 22 and 11 $\sigma$ respectively.

Long-term data from \rxte/\textit{ASM} shows wide variations in the intensity of \src. Similar studies performed with \gsc\ showed that \src\ exhibits multiple luminosity levels with no apparent super-orbital period~\citep{Ajith_cenx3_2024}. From the long-term \rxte/\textit{ASM} lightcurve, we can see that during this observation, the source was in a high intensity level (Fig.~\ref{fig:rxte_asm_long_term}). \xray\ luminosity states in \src\ has been studied in the past, and several hypotheses have been put forward to explain the multiple source intensities. The source was captured in a transition from low intensity to high intensity level with \chandra\ telescope~\citep{Sanjurfo_ferrin_CenX3_Chandra_2024}. The transition was hypothesised to be due to the cooling of a large chunk of matter, which increased the mass accretion onto the NS~\citep{Sanjurfo_ferrin_CenX3_Chandra_2024}. From \rxte\ observations of \src\ it was proposed that a aperiodically precessing warped accretion disk is probably responsible for the changes in intensity states of \src~\citep{Raichur_CenX3_long_term_intensity_2008}. However, a later study of \src\ with \astrosat\ showed that spectral shape and pulse profile are different in different luminosity states, pointing to accretion column origin~\citep{Bachhar_CenX3_AstroSat_2022}.

\begin{figure*}
    \centering
    \includegraphics[width=0.99\linewidth]{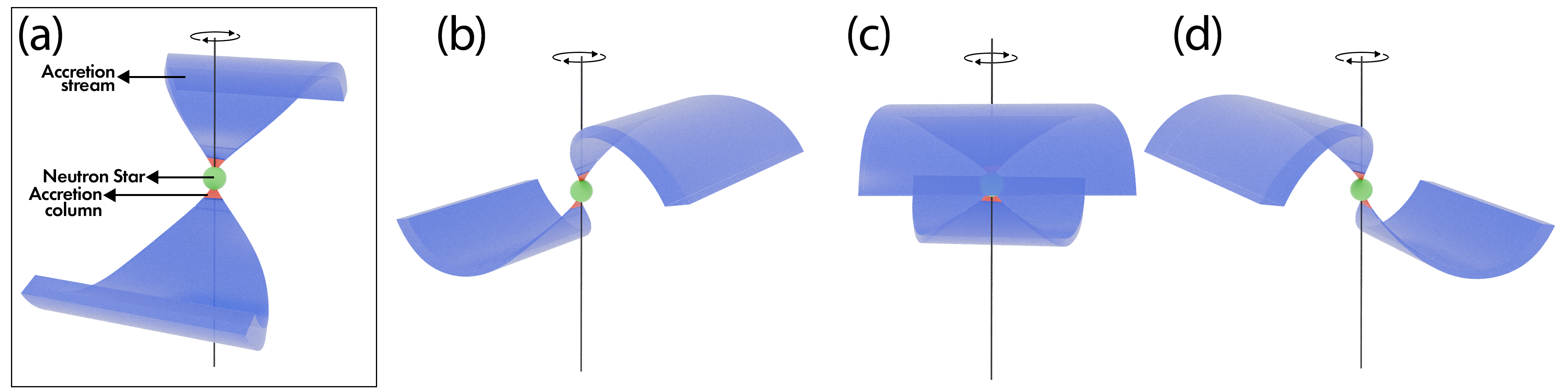} 
    \caption{The figure shows a schematic representation of \src. The compact object (green), accretion column (red) and the accretion stream (blue) are shown in different colours in Panel (a). In Panel b-d, we see a different line of sight to the NS. Panel c shows when the accretion stream intercepts our line of sight to the NS.}
    \label{fig:Cen X3 orientation}
\end{figure*}

Emissions from different ionisation states of iron probe different regions around the NS. Intensity and equivalent width variations of the three iron K lines with orbital phase of \src\ were investigated with a \suzaku\ observation~\citep{Naik_cenx3_suzaku_2011}. 
Using \asca~\citep{Ebisawa_cenx3_asca_1996} and \xmm~\citep{Naik_cenx3_xmm_2012} observations, it was shown that the 6.4 keV fluorescence lines may originate from close to the NS, while the lines at 6.7 and 6.97 keV are produced further away, either in the photo-ionised wind of the MS star or in the accretion disk corona. The high resolving power of \xrism\ allows investigation of the Doppler motion of the 6.4 keV iron \kalpha\ emission line. From the measurements of radial velocity around the binary orbit, the source of the fluorescent line could not be attributed to either the NS surface, the companion stellar wind, the companion star surface, or the accretion wake alone~\citep{Mochizuki_CenX3_XRISM_2024}. Elaborate models of iron fluorescence is required for the complete understanding of the origin of the emission line. 

In this work, we confirmed the pulse phase variation of the three iron \kalpha\ lines and report the pulse phase variation observed in the iron \kbeta\ line. We investigated the reprocessing region in \src\ using the pulse phase dependency of the iron emission lines. The variations of the four emission lines with pulse phase are plotted in Fig.~\ref{fig:PRS_line_strength_pulse_profile}. All four lines show significant variability with the pulse phase of the NS. The pulse phase variability of the Fe I \kalpha, Fe XXV \Healpha, Fe XXVI \Lyalpha\, and Fe \kbeta\ lines were detected at 8.8, 8.2, 5.5 and 18.9 $\sigma$. The modulations of the four iron lines are single-peaked in nature, unlike the pulse profile, which has two peaks. The pulse phase variability of the iron fluorescent lines could be well fitted with a sinusoidal function on top of a constant value. The modulation in the iron line with pulse phase was quantified by taking the ratio of the sine amplitude to the constant fit. The modulation factor of the Fe XXVI \Lyalpha\ ( 31.0 $\pm$ 7.8 \%) is much stronger than the Fe XXV \Healpha\ ( 18.1 $\pm$ 3.0 \%) and the near-neutral iron \kalpha\ ( 13.1 $\pm$ 1.6 \%) emission line. The modulation factor of the \kbeta\ line was calculated to be 50.5 $\pm$ 5.6\%. The pulse phase corresponding to the minima of pulse phase dependence for the four emission lines coincides (Section~\ref{sec:PRS}), indicating a common origin.

The evolution of the line intensity during eclipse egress in \src\ shows that the highly ionised line species has a smaller ``rise time" and hence it originates closer to the NS ~\citep{Ebisawa_cenx3_asca_1996, Naik_cenx3_xmm_2012} while the neutral iron \kalpha\ emission line takes about 40 ks to reach its maximum value as \src\ comes out of eclipse~\citep{Naik_cenx3_xmm_2012}. The neutral line-forming region probably encompasses a region close to 55 lt-s in size. A similar calculation of the line forming region with \asca\ data showed a size of $\sim$ 28 lt-s~\citep{Ebisawa_cenx3_asca_1996}. This region is probably responsible for the non-pulsating component. However, the region emitting the pulsed component of the iron line should be much more compact and within 4.8 lt-s as the pulse period of the NS is only about 4.8 s.
The pulse phase-dependent modulation in the four iron emission lines indicates the existence of a reprocessing region close to the NS. 

Naively, an anisotropic distribution of matter producing the pulse phase variability can be from clumps in the stellar wind, which is expected from the massive O-type companion star in \src. However, the identical pulse phase modulation of the four iron fluorescent lines across a wide orbital phase range (Section~\ref{sec:TR_PRS}) shows no significant change, indicating that the reprocessing emission must come from structures that are much more persistent than clumpy wind material. The absence of any change of the relative phase between modulation of the iron lines and the pulse profile with the orbital phase of \src\ indicates that the component responsible for the pulsed emission must have very little relative motion with respect to the NS and the observer. The absorption column density and the strength of the iron line show similar variations with pulse phase~\citep{Day_CenX3_pulsed_Fe_1993}. The absorption column density as well as the emission line strength in \src\ is maximum between the two peaks in the pulse profile of \src~\citep{Day_CenX3_pulsed_Fe_1993}.

In the case accretion by Roche Lobe overflow happens in \src, then at the location where accreted matter from the secondary star meets the accretion disc in \src\ can be the origin of the pulsed component of the iron emission lines. In this scenario, the pulse phase modulations from the multiple iron emission lines are expected to be in phase. The location of the hotspot, however, is expected to vary with respect to our line of sight to the neutron star in different orbital phases. The companion star surface sustains a large solid angle with respect to the NS and therefore, it is not expected to produce a very sharp pulse phase modulation in the emission component. Another possible origin of the anisotropy can be from the wake-like structure observed in \src~\citep{Ajith_cenx3_2024}, however, this structure is also expected to show orbital variation in the pulse-phase modulations of the multiple iron lines. In the current observation, we see that the iron lines do not show significant variability with phase, rejecting this scenario as the origin of the pulsed emission lines.

We propose a possible physical configuration to explain the observed pulse phase variation of iron fluorescence lines in \src. A sketch of the NS system is given in Fig.~\ref{fig:Cen X3 orientation}(a) similar to Fig. 10 of \cite{Tsygankov_cenx3_ixpe_2022}. The NS is shown in green, the accretion column in red and the accretion stream channelling matter from the accretion disc to the NS is shown in blue. The accretion stream corotates with the NS spin motion as it is attached to the magnetic field of the compact object. \xray\ from the NS are emitted from the accretion column at the base of the accretion stream. Assuming the accretion column forms a broad beam, the accretion stream partially obstructs the view of the \xray\ from the accretion column as the base of the accretion stream, where the accretion stream meets the accretion disc, comes into our line of sight (Fig.~\ref{fig:Cen X3 orientation} b$-$d). The presence of the accretion stream produce two humps in the pulse profile with an increase of absorption column density in between the two peaks. At the same orientation of the NS, the base of the accretion stream (Fig.~\ref{fig:Cen X3 orientation} c), would also produce an increase in fluorescent emission. Such an increase in absorption column density and iron line strength between the two peaks of the pulse profile has been observed in \src\ with \ginga\ data~\citep{Day_CenX3_pulsed_Fe_1993}. We see a similar pattern in the pulse phase dependence of the four iron fluorescence strength and pulse profile (Fig.~\ref{fig:PRS_line_strength_pulse_profile}) consistent with \ginga\ results. At any orbital phase, the pulse phase variations of the iron lines are expected to have the same phase relation with the pulse profile, i.e., the maximum intensity of the lines occurring at the dip between the two pulse peaks.
The similarity of the shape of the pulse variation of the four iron line (Section~\ref{sec:PRS}) and their persistence with orbital motion (Section~\ref{sec:TR_PRS}) of \src\ can be reasonably explained by the above scenario. Therefore, the accretion stream cutting the line of sight to \src\ can produce the observed pulse phase dependence of the multiple iron lines.

\section{Summary} \label{sec:summary}

In this work, we analyse an \xmm\ observation of \src\ to study the binary reprocessing environment. Multiple prominent lines were present in the source spectrum, and we have carried out pulse phase resolved spectroscopy to investigate the three \kalpha\ lines from near neutral, Helium-like and Hydrogen-like iron and the iron \kbeta\ emission lines. The results from the phase-resolved analysis can be summarised as follows :

\begin{itemize}
    \item The near-neutral Fe I \kalpha, Fe XXV \Healpha, Fe XXVI \Lyalpha\ and Fe \kbeta\ lines are variable with the pulse phase of the NS.
    \item The Fe XXVI \Lyalpha\ (31\%) and Fe \kbeta\ (51\%) showed much stronger modulation than the Fe XXV \Healpha\ (18 \%) and the Fe I \kalpha\ (13\%) emission line.
    \item The pulsating and non-pulsating iron fluorescent emission probably originates from different locations in the binary system.
    \item The shape of the pulse phase modulations appears in phase and does not show significant variation with the orbital motion of the NS.
    \item The pulsating component for the four iron fluorescent lines probably shares a common origin and must originate within the light travel distance corresponding to the pulse period of the NS.
    \item The pulsating component can originate from an accretion stream intersecting the line of sight to the accretion column of the NS from the observer.
\end{itemize}

\vspace{5mm}



\section*{Acknowledgments}
We thank the referee for the useful comments and suggestions that improved the quality of this paper. KR would like to thank Gulab Dewangan and Matteo Guainazzi for the discussion regarding instrumental effects in EPIC$-$PN spectrum. KR would like to thank Arunima for creating the illustrations.



\bibliography{CenX3_XMM}{}
\bibliographystyle{aasjournal}



\end{document}